\newcommand{\re}{\mathrm{e}}            
\documentclass[12pt]{iopart}
\begin{document}

\def\bm#1{\mbox{\boldmath $#1$}}
\def\s{\sigma}
\def\I{\rm i}
\newcommand{\be}{\begin{eqnarray}}
\newcommand{\ee}{\end{eqnarray}}
\newcommand{\no}{\nonumber}

\input epsf.sty

\title[Random walks with absorbing boundaries]{Exact solution 
for random walks on the triangular lattice with absorbing boundaries}

\author{M T Batchelor\dag\ and B I Henry\ddag} 

\address{\dag\ Department of Theoretical Physics, Research School of Physical
Sciences and Engineering, and Centre for Mathematics and its Applications,
School of Mathematical Sciences, 
The Australian National University, Canberra ACT 0200, Australia}

\address{\ddag\ Department of Applied Mathematics, School
of Mathematics, University of New South Wales,
Sydney NSW 2052, Australia}

\begin{abstract}
The problem of a random walk on a finite triangular lattice
with a single interior source point and zig-zag absorbing boundaries is solved
exactly. This problem has been previously considered intractable.
\end{abstract}




\section{Introduction}

The problem of a random walk on a two-dimensional lattice
with a single interior source point and absorbing boundaries
was first considered by Courant \textit{et al.} \cite{C28} in 1928
where general properties of the solution were discussed.
This problem was solved exactly in
1940 \cite{MW40} for the case of
random walks on a square lattice with rectangular
absorbing boundaries.
The problem on a triangular lattice with finite absorbing boundaries
was considered in 1963 \cite{KM63} where the
exact solution was given for an approximation of the problem
using straight boundaries rather than the true
zig-zag boundaries of the triangular lattice.
Indeed the authors of this paper remarked that ``An explicit solution of 
the difference equation can hardly be obtained if these boundary 
conditions are used.''
Other variants of the problem on the square lattice
have been solved exactly \cite{KM63,M94}
however the problem on the triangular lattice
with true zig-zag triangular lattice boundaries
has remained unsolved.
In this paper we give the exact solution for this
problem.

The problem of random walks on finite lattices with absorbing
boundaries is fundamental to the
theory of stochastic processes \cite{ML79} and has numerous applications.
These include
potential theory \cite{D53},
electrical networks \cite{DS84},
surface diffusion \cite{BCMO99} and Diffusion-Limited Aggregation
(DLA)
\cite{WS81}. For example,
the exact results for the square lattice problem
\cite{MW40} have been used to
expedite the growth of large DLA
clusters on the square lattice \cite{BHR95}
and the formulae derived in this paper
could similarly
be used on the triangular lattice.

\section{Field equations, boundary conditions and absorption probabilities}

A schematic illustration showing random walk pathways (dashed lines) on an equi-
angular triangular lattice is shown in Fig. 1.
A natural co-ordinate system for the triangular lattice
is to label the lattice vertices by the intersection points
$(p,q)$ of horizontal straight lines
$p=0,1,2,\ldots m+1$
and slanted straight lines $q=0,1,2,\ldots n+1$ -- parallel to the
$p$ axis shown in Fig. 1.
In this co-ordinate system the expectation that a random
walk on the lattice visits a site $(p,q)$,
before exiting at a finite boundary, is given by the non-separable
difference equation
\begin{eqnarray*}
F(p,q)&=&\frac{1}{6}\left[F(p,q-1)+F(p,q+1)+F(p-1,q)\right.\\
& &\quad \left. +F(p+1,q)+F(p+1,q-1) +F(p-1,q+1)\right].
\end{eqnarray*}
In this paper we adopt a different co-ordinate system, which may
at first appear less natural but it has the advantage that
the governing equations are separable.

\begin{figure}[h!]
\vspace{80mm}
\includegraphics{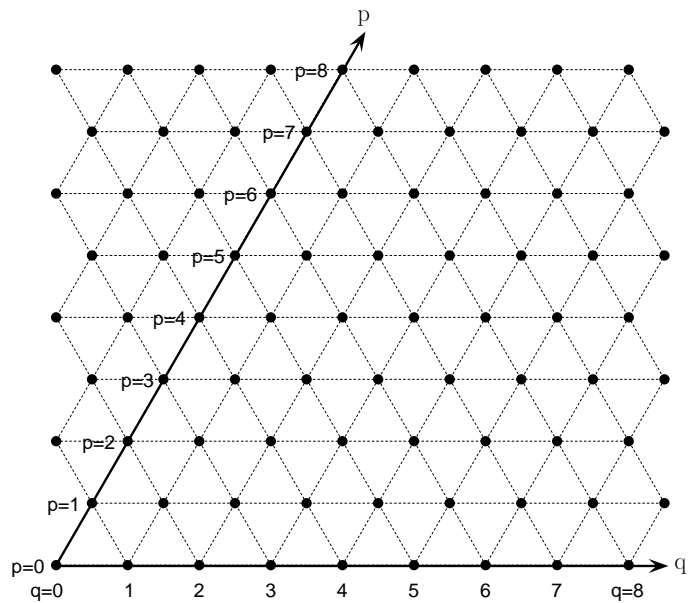}
\caption{An equi-angular triangular lattice with
natural $(p,q)$ co-ordinates.}
\end{figure}

In the co-ordinate system used here
we label the vertices of the  triangular lattice
by the intersection points $(p,q)$ of horizontal straight lines
$p=0,1,2,\ldots m+1$
and vertical zig-zag lines $q=0,1,2,\ldots n+1$ (see Fig. 2).
In the absence of any source terms the expectation
that a random walker visits a lattice site $(p,q)$ is given by
the coupled homogeneous difference equations
\begin{eqnarray}
F(p,q)&=&\frac{1}{6}\left[\hat F(p-1,q-1)+F(p,q-1)+\hat F(p+1,q-1)\right.\nonumber\\
& &\quad \left. +\hat F(p+1,q) +F(p,q+1)+\hat F(p-1,q)\right],\label{Feq}\\
\hat F(p,q)&=&\frac{1}{6}\left[ F(p-1,q)+\hat F(p,q-1)+F(p+1,q)\right.\nonumber\\
& &\quad \left. + F(p+1,q+1) +\hat F(p,q+1)+ F(p-1,q+1)\right],\label{Fhateq}
\end{eqnarray}
where $F(p,q)$ denotes the field value at an even $p$ co-ordinate and
$\hat F(p,q)$ denotes the field value at an odd $p$ co-ordinate.
This separation of even and odd field equations,
which is not necessary for the problem on the
square lattice,
 is central to our
solution below.

\begin{figure}[h!]
\vspace{80mm}
\includegraphics{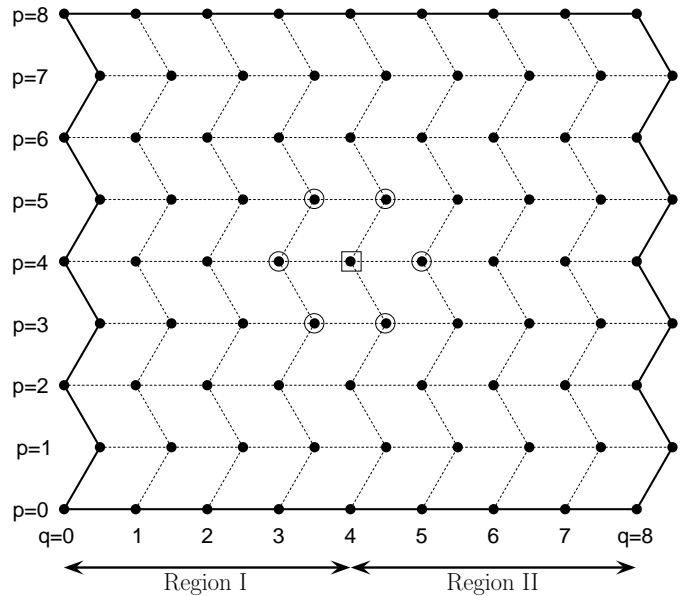}
\caption{Co-ordinate system, represented by dashed lines,
adopted in this paper for random walks
on the triangular lattice with true zig-zag lattice boundaries.
In this example $m=7, n=7$. Nearest neighbour sites, highlighted
with open circles, are shown surrounding a source point, highlighted by
an open box, at $a=4, b=4$. The absorbing boundary sites are connected by solid
lines.}
\end{figure}

Now consider a source term at $(a,b)$ and absorbing boundaries at
$p=0,m+1$ and $q=0,n+1$. In Section 3 we give the solution for the case
where $m>1$ is odd. The same methods can be used
with slightly altered equations to obtain the solution
for $m$ even. The special case of the strip with $m=1$ has been
detailed elsewhere \cite{BH02}.
Following McCrea and  Whipple \cite{MW40} we construct separate solutions,
$F_I(p,q),  \hat F_I(p,q)$
 for
$q\le b$ and $F_{II}(p,q), \hat F_{II}(p,q)$ for $q\ge b$.
The absorbing boundary conditions are thus given by
\begin{eqnarray}
F(0,q)&=&0,\\
F(m+1,q)&=&0,\\
F_I(p,0)&=&0,\\
\hat F_I(p,0)&=&0,\\
F_{II}(p,n+1)&=&0,\\
\hat F_{II}(p,n+1)&=&0.
\end{eqnarray}
The ommision of a subscript $I$ or $II$  in the above
indicates that the same equations are satisfied by
both
$F_I$ and $F_{II}$.

At $q=b$ we have the matching conditions
\begin{eqnarray}
F_I(p,b)&=&F_{II}(p,b),\\
\hat F_I(p,b)&=&\hat F_{II}(p,b).
\end{eqnarray}
Finally, with the inclusion of the source term, we have the
coupled inhomogeneous field equations at $q=b$,
\begin{eqnarray}
6F_I(p,b)&=&6\delta_{p,a}((a+1) \mbox{mod} 2) \nonumber\\
& &+\left[\hat F_I(p-1,b-1)+F_I(p,b-1)+\hat F_I(p+1,b-1)\right.\nonumber\\
& &\left. \quad + \hat F_I(p+1,b)+F_{II}(p,b+1)+\hat F_I(p-1,b)\right],\\
6\hat F_I(p,b)&=&6\delta_{p,a}(a \mbox{mod} 2)\nonumber\\
& &+\left[F_I(p-1,b)+\hat F_I(p,b-1)+F_I(p+1,b)\right.\nonumber\\
& &\left. \quad +F_{II}(p+1,b+1) +\hat F_{II}(p,b+1)+F_{II}(p-1,b+1)\right].
\end{eqnarray}
In Section III we present the solutions to Eqs. (1)--(12).

The probabilities for the random walking particle
to be absorbed at any
specified point on one of the four
boundaries,
$p=0, p=m+1, q=0, q=n+1$, are readily obtained by averaging over the
nearest neighbour expectation values subject to the boundary conditions,
Eqs. (3)--(8).
For example,
\begin{eqnarray}
P(2k-1,0)&=&\frac{1}{6}\left[\hat F_I(2k-1,1)+F_I(2k-2,1)
+F_I(2k,1)\right];\nonumber\\
& &\qquad\quad k=1,\ldots \frac{m+1}{2}.
\end{eqnarray}

\section{Solution}

\subsection{Homogeneous equations -- Separation of variables}

We begin by solving the homogenous equations, Eqs. (1),(2),
subject to the absorbing boundary conditions, Eqs. (3),(4).
The homogenous field equations separate, on using
\begin{eqnarray}
F(p,q)&=&P(p)Q(q),\label{sepFeq}\\
\hat F(p,q)&=&\hat P(p)\hat Q(q)\label{sepFhateq},
\end{eqnarray}
into
\begin{eqnarray}
\frac{6Q(q)-Q(q+1)-Q(q-1)}{\hat Q(q-1)+\hat Q(q)}&=&
\frac{\hat P(p-1)+\hat P(p+1)}{P(p)}=\lambda,\\
& &\nonumber\\
\frac{6\hat Q(q)-\hat Q(q+1)-\hat Q(q-1)}{Q(q+1)+ Q(q)}&=&
\frac{P(p-1)+ P(p+1)}{\hat P(p)} =\kappa,
\end{eqnarray}
where $\kappa$ and $\lambda$ are separation constants.
The two coupled equations for $P$ and $\hat P$ can be readily
 decoupled into separate
equations
\begin{eqnarray}
P(2k+2)+(2-\lambda\kappa)P(2k)+P(2k-2)=0,&&\\
\hat P(2k+3)+(2-\lambda\kappa)\hat P(2k+1)+\hat P(2k-1)=0,&&
\end{eqnarray}
where $k$ is an integer. This
second order linear difference equation
for $P, (\hat P)$ on a lattice of
even (odd) integers has the solution
$$
P(2k)=A\mu^k+B\mu^{-k},
$$
where
$$
\mu=\frac{-(2-\lambda\kappa)+\sqrt{(2-\lambda\kappa)^2-4}}{2},
$$
and $A$ and $B$ are arbitrary.
It follows that the solution of $P(p)$ (with $p$ even) corresponding to the
boundary conditions, Eqs. (3),(4), is
\begin{equation}
P(p)=c\sin(\frac{\pi j p}{m+1})\label{Psol}
\end{equation}
with
\begin{equation}
\lambda\kappa=4\cos^2(\frac{\pi j}{m+1}).
\end{equation}
The solution for $\hat P(p)$ (with $p$ odd) now follows from Eq. (17) as
\begin{equation}
\hat P(p)=\frac{2c}{\kappa}\cos(\frac{\pi j}{m+1})\sin(\frac{\pi j p}{m+1}).
\label{Phatsol}\end{equation}

Equations (16) and (17) can also be decoupled for $Q(q)$ and $\hat Q(q)$
resulting in the same fourth order linear difference equation in each case;
\begin{eqnarray}
Q(q+4)-(12+\lambda\kappa)Q(q+3)+(38-2\lambda\kappa)Q(q+2)&&\nonumber\\
\qquad\qquad -(12+\lambda\kappa)Q(q+1)+Q(q)=0.&&
\end{eqnarray}
The zeroes of the corresponding characteristic quartic polynomial
are in reciprocal pairs  leading to the
solutions
\begin{eqnarray}
Q(q)&=&c_1\re^{\alpha q}+c_2\re^{-\alpha q}+c_3\re^{\beta q}+c_4\re^{-\beta q},
\label{Qeq}\\
\hat Q(q)&=&\hat c_1\re^{\alpha q}+\hat c_2\re^{-\alpha q}+\hat c_3\re^{\beta q}
+\hat c_4\re^{-\beta q},\label{Qhateq}
\end{eqnarray}
where
\begin{eqnarray}
\cosh\alpha&=&3+\frac{\lambda\kappa}{4}+\frac{1}{4}\sqrt{(\lambda\kappa)^2
+32\lambda\kappa},\\
\cosh\beta&=&3+\frac{\lambda\kappa}{4}-\frac{1}{4}\sqrt{(\lambda\kappa)^2
+32\lambda\kappa}.
\end{eqnarray}
The coefficients $c_1, c_2, c_3, c_4$ may be chosen arbitrarily
with
the coefficients $\hat c_1, \hat c_2, \hat c_3, \hat c_4$ then determined
by
substituting Eqs. (\ref{Qeq}), (\ref{Qhateq}) into Eq. (16)
or Eq. (17).
{}From Eq. (17) we obtain
\begin{equation}
\hat c_1=c_1 f(\alpha),\, \hat c_2=c_2 f(-\alpha),\, \hat c_3=c_3 f(\beta),\,
\hat c_4=c_4 f(-\beta)
\end{equation}
where
\begin{equation}
f(\omega)=\frac{\kappa (\re^{\omega}+1)}{2(3-\cosh\omega)}.
\end{equation}

By combining Eqs. (14),(15),(20),(22),(24),(25),(28),(29)
 we can write the solutions to the
field equations, Eqs. (\ref{Feq}), (\ref{Fhateq}), that satisfy
 the boundary conditions, Eqs. (3),(4), in the form
\begin{eqnarray}
& &F(p,q;A,B,C,D)=\sum_{j=1}^m\sin\left(\frac{\pi j p}{m+1}\right)\nonumber\\
& &\quad\times
\left[A_j\re^{\alpha q}(3-\cosh\alpha)
+B_j\re^{-\alpha q}(3-\cosh\alpha)\right.\nonumber\\
& &\qquad\left.+ C_j\re^{\beta q}(3-\cosh\beta)+D_j\re^{-\beta q}(3-\cosh\beta)
\right],\\
& &\nonumber\\
& &\hat F(p,q;A,B,C,D)=\sum_{j=1}^m\sin\left(\frac{\pi j p}{m+1}\right)
\cos\left(\frac{\pi j}{m+1}\right)\nonumber\\
& &\quad\times\left[
A_j\re^{\alpha q}(\re^\alpha+1)
+ B_j\re^{-\alpha q}(\re^{-\alpha} +1)\right.\nonumber\\
& &\qquad\left. + C_j\re^{\beta q}(\re^\beta+1)+
D_j\re^{-\beta q}(\re^{-\beta}+1)\right],
\end{eqnarray}
where $A_j,B_j,C_j,D_j$
are arbitrary.

Before imposing the remaining absorbing boundary conditions,
Eqs. (5)--(8),
we represent the solutions in the two regions: $I: q\le b$; $II: q\ge b$ by
\begin{eqnarray*}
F_I(p,q)&=&F(p,q;A,B,C,D),\\
\hat F_I(p,q)&=&\hat F(p,q;A,B,C,D),\\
F_{II}(p,q)&=&F(p,q;\tilde A,\tilde B,\tilde C,\tilde D),\\
\hat F_{II}(p,q)&=&\hat F(p,q;\tilde A,\tilde B,\tilde C,\tilde D).
\end{eqnarray*}
The absorbing boundary
conditions, Eqs. (5)--(8), can be used
to eliminate
$C,D,\tilde C,\tilde D$. After an appropriate re-normalization of the
coefficients $A,B,\tilde A, \tilde B$ this yields
\begin{eqnarray}
F_I(p,q)&=&\sum_{j=1}^m\sin\left(\frac{\pi j p}{m+1}\right)\nonumber\\
& &\times\left[A_I(q,\alpha,\beta)A_j +A_I(q,-\alpha,\beta)B_j\right],\label{FI}
\\
& &\nonumber\\
\hat F_I(p,q)&=&\sum_{j=1}^m\sin\left(\frac{\pi j p}{m+1}\right)
\cos\left(\frac{\pi j}{m+1}\right)\nonumber\\
& &\times\left[\hat A_I(q,\alpha,\beta)A_j+\hat A_I(q,-\alpha,\beta)B_j\right],
\label{FIhat}\\
& &\nonumber\\
F_{II}(p,q)&=&\sum_{j=1}^m\sin\left(\frac{\pi j p}{m+1}\right)\nonumber\\
& &\times\left[A_{II}(q,\alpha,\beta)\tilde A_j+A_{II}(q,-\alpha,\beta)
\tilde B_j\right],\label{FII}\\
& &\nonumber\\
\hat F_{II}(p,q)&=&\sum_{j=1}^m\sin\left(\frac{\pi j p}{m+1}\right)
\cos\left(\frac{\pi j}{m+1}\right)\nonumber\\
& &\times\left[\hat A_{II}(q,\alpha,\beta)\tilde A_j+
\hat A_{II}(q,-\alpha,\beta)\tilde B_j\right],\label{FIIhat}
\end{eqnarray}
where
\begin{eqnarray}
A_I(q,\alpha,\beta)&=&(3-\cosh\alpha)(3-\cosh\beta)2\sinh\beta \re^{\alpha q}
\nonumber\\
&& -(3-\cosh\beta)\gamma(-\alpha,-\beta)\re^{\beta q}\nonumber\\
&& +(3-\cosh\beta)\gamma(-\alpha,\beta)\re^{-\beta q},\label{AI}\\
& &\nonumber\\
\hat A_I(q,\alpha,\beta)&=&2\re^{\alpha q}(\re^\alpha+1)(3-\cosh\beta)\sinh\beta
\nonumber\\
&& -\gamma(-\alpha,-\beta)(\re^\beta+1)\re^{\beta q}
 +\gamma(-\alpha,\beta)(\re^{-\beta}+1)\re^{-\beta q},\label{AIhat}\\
& &\nonumber\\
A_{II}(q,\alpha,\beta)&=&(3-\cosh\alpha)(3-\cosh\beta)2\sinh\beta \re^{\alpha q}
\nonumber\\
&& -\re^{(\alpha-\beta)(n+1)}\gamma(-\alpha,-\beta)(3-\cosh\beta)\re^{\beta q}
\nonumber\\
&& +\re^{(\alpha+\beta)(n+1)}\gamma(-\alpha,\beta)(3-\cosh\beta)\re^{-\beta q},
\label{AII}\\
& &\nonumber\\
\hat A_{II}(q,\alpha,\beta)&=&(\re^\alpha+1)(3-\cosh\beta)2\sinh\beta 
\re^{\alpha q}\nonumber\\
&& -\re^{(\alpha-\beta)(n+1)}\gamma(-\alpha,-\beta)(\re^\beta+1)\re^{\beta q}
\nonumber\\
&& +\re^{(\alpha+\beta)(n+1)}\gamma(-\alpha,\beta)(\re^{-\beta}+1)
\re^{-\beta q},
\label{AIIhat}
\end{eqnarray}
and the function
\begin{eqnarray}
\gamma(\alpha,\beta)&=&4\cosh\alpha-4\cosh\beta-(3-\cosh\beta)
\sinh\alpha\nonumber\\
& &\quad-(3-\cosh\alpha)\sinh\beta\label{gamma}.
\end{eqnarray}
The coefficients defined in Eq. (36)
which are shown as implicit functions of
$\alpha$ and $\beta$
are also functions of $j$ via Eqs. (21),(26),(27).
Equations (\ref{FI})--(\ref{FIIhat}),
are general solutions to the coupled homogeneous field equations,
Eqs. (1),(2),
that satisfy all of the absorbing boundary conditions, Eqs. (3)--(8).

\subsection{Inhomogeneous equations -- Matching conditions}
By using the two matching conditions, Eqs. (9),(10),
the four arbitrary constants, $A, B, \tilde A, \tilde B$,
can be reduced to two arbitrary constants, $A,B$ say.
The solutions in Region II
can then be written as
\begin{eqnarray}
F_{II}(p,q)&=&\sum_{j=1}^m\sin\left(\frac{\pi j p}{m+1}\right)
\left[A_{II}^\star(q,\alpha,\beta)A_j+A_{II}^\star(q,-\alpha,\beta)B_j \right],\\
& &\nonumber\\
\hat F_{II}(p,q)&=&\sum_{j=1}^m\sin\left(\frac{\pi j p}{m+1}\right)
\cos\left(\frac{\pi j}{m+1}\right)\nonumber\\
& &\times\left[\hat A_{II}^\star(q,\alpha,\beta)A_j +\hat A_{II}^\star(q,-\alpha
,\beta)B_j\right],
\end{eqnarray}
where
\begin{eqnarray}
A_{II}^\star(q,\alpha,\beta)=A_{II}(q,\alpha,\beta)\Gamma_1+A_{II}(q,-\alpha,\beta)
\Gamma_2,&&\\
&&\nonumber\\
\hat A_{II}^\star(q,\alpha,\beta)=\hat A_{II}(q,\alpha,\beta)\Gamma_1+\hat A_{II
}(q,-\alpha,\beta)\Gamma_2,&&
\end{eqnarray}
and
\begin{eqnarray}
\Gamma_1=\frac{\hat A_{II}(b,-\alpha,\beta)A_I(b,\alpha,\beta)-A_{II}(b,-\alpha,
\beta)\hat A_I(b,\alpha,\beta)}
{A_{II}(b,\alpha,\beta)\hat A_{II}(b,-\alpha,\beta)-\hat A_{II}(b,\alpha,\beta)
A_{II}(b,-\alpha,\beta)},&&\\
&&\nonumber\\
\Gamma_2=\frac{\hat A_{II}(b,\alpha,\beta)A_I(b,\alpha,\beta)-A_{II}(b,\alpha,
\beta)\hat A_I(b,\alpha,\beta)}
{A_{II}(b,-\alpha,\beta)\hat A_{II}(b,\alpha,\beta)-A_{II}(b,\alpha,\beta)
\hat A_{II}(b,-\alpha,\beta)},&&
\end{eqnarray}

Finally the remaining arbitrary constants
$A,B$ are determined from the requirement that the solutions
satisfy the coupled inhomogeneous field equations, Eqs. (11),(12).
This step is facilitated using the identity
\begin{equation}
\delta_{p,a}=\frac{2}{m+1}\sum_{j=1}^m\sin\left(\frac{\pi j a}{m+1}\right)
\sin\left(\frac{\pi j p}{m+1}\right).
\end{equation}
Explicitly we find:\\
i) $a$ even,
\begin{eqnarray}
&&A_j=\frac{T_1(j)}{T_2(j)}B_j,\\
&&B_j=\frac{12}{m+1}\sin\left(\frac{\pi j a}{m+1}\right)\frac{T_2(j)}{T_1(j)T_3(
j)-T_4(j)T_2(j)};
\end{eqnarray}
ii) $a$ odd,
\begin{eqnarray}
&&A_j=\frac{T_4(j)}{T_3(j)}B_j,\\
&&B_j=\frac{12}{m+1}\sin\left(\frac{\pi j a}{m+1}\right)\frac{T_3(j)}{T_2(j)T_4(
j)-T_1(j)T_3(j)},
\end{eqnarray}
and
\begin{eqnarray}
T_1(j)&=&2A_I(b,-\alpha,\beta)+2A_{II}^\star(b+1,-\alpha,\beta)
+\hat A_I(b-1,-\alpha,\beta)\nonumber\\
& &+\hat A_{II}^\star(b+1,-\alpha,\beta)
 -6\hat A_I(b,-\alpha,\beta),\\
& &\nonumber\\
T_2(j)&=&
6\hat A_I(b,\alpha,\beta)-2A_I(b,\alpha,\beta)-2A_{II}^\star(b+1,\alpha,\beta)
\nonumber\\
& &-\hat A_I(b-1,\alpha,\beta)-\hat A_{II}^\star(b+1,\alpha,\beta),\\
& &\nonumber\\
T_3(j)&=&6A_I(j,b)-2\cos^2(\frac{\pi j}{m+1})\hat A_I(b-1,\alpha,\beta)\nonumber
\\
& &-2\cos^2(\frac{\pi j}{m+1})\hat A_I(b,\alpha,\beta)-A_I(b-1,\alpha,\beta)
\nonumber\\
& &-A_{II}^\star(b+1,\alpha,\beta),\\
& &\nonumber\\
T_4(j)&=&2\cos^2(\frac{\pi j}{m+1})\hat A_I(b-1,-\alpha,\beta)\nonumber\\
& &+2\cos^2(\frac{\pi j}{m+1})\hat A_I(b,-\alpha,\beta)
+A_I(b-1,-\alpha,\beta)\nonumber\\
& &+A_{II}^\star(b+1,-\alpha,\beta)-6A_I(b,-\alpha,\beta).
\end{eqnarray}

Our solution for the triangular lattice site
expectation values with absorbing boundary conditions and a
source point is finally given by
Eqs. (32),(33) in Region I and
Eqs. (41),(42) in Region II. The relevant quantities
appearing in these equations are defined through the series
of equations, Eqs. (21),(26),(27),(36)--(40),(43)--(46),(48)--(50).

\section{Example}
Consider the case of
a triangular lattice with $m=7, n=7$ and a source at $a=4, b=4$
(Fig. 2).
In the approximation to this problem using straight edge
 boundaries and the
$(p,q)$ co-ordinate system of
Keberle and Montet \cite{KM63} this problem corresponds to;
$m=16, n=7, a=8, b=4$.

We have calculated expectation values at the nearest neighbour
lattice sites around the source and absorption
probabilities at the boundaries; (a) using our exact results above,
and (b) using
the results of Keberle and Montet \cite{KM63} for the approximation
to the problem -- their equations (7a), (7b), (7c).
In this small model lattice system we found that
the straight line boundary
approximation provides reasonable results for expectation values
near the source (accurate to within a few percent) but provides
poor results
for the absorption probabilities (Table 1).
Note that the `absorption probabilities'
for the zig-zag boundary co-ordinates which are inside the straight line
boundary of Keberle and Montet \cite{KM63} are not true probabilities
in their solution.

\begin{table}
\caption{Absorption probabilities at the boundaries
on a triangular lattice with a source point enclosed by zig-zag lattice
 boundaries; (a)
exact results using the formulae derived in this paper,
(b) results using equations (1), (7a), (7b), (7c)
derived in Keberle and Montet \cite{KM63} for the
straight line boundary approximation.\label{tabone}}

\begin{indented}
\lineup
\item[]\begin{tabular}{@{}*{5}{l}}
\br
$\0\0$(a)&&&(b)&\cr
\mr
$P(0,1)$&.012247&&$P(2,0)$&.019609\\
$P(0,2)$&.035646&&$P(4,0)$&.039949\\
$P(0,3)$&.054827&&$P(6,0)$&.057904\\
$P(0,4)$&.063296&&$P(8,0)$&.065834\\
$P(0,5)$&.056686&&$P(10,0)$&.059209\\
$P(0,6)$&.039811&&$P(12,0)$&.042817\\
$P(0,7)$&.020737&&$P(14,0)$&.024583\\
$P(0,8)$&.005743&&$P(16,0)$&.007903\\
$P(1,0)$&.026040&&$P(1,1)$&.033624\\
$P(2,0)$&.013797&&$P(0,2)$&.036214\\
$P(3,0)$&.069523&&$P(1,3)$&.088151\\
$P(4,0)$&.021476&&$P(0,4)$&.053888\\
$P(1,8)$&.005743&&$P(17,1)$&.015276\\
$P(2,8)$&.040253&&$P(16,2)$&.051612\\
$P(3,8)$&.015039&&$P(17,3)$&.038562\\
$P(4,8)$&.059725&&$P(16,4)$&.075927\\
\br
\end{tabular}
\end{indented}
\end{table}

\section{Conclusion}

Although our solution is somewhat unwieldly, we have nevertheless solved
the underlying field equations for random walks on a finite triangular lattice
with a single interior source point and zig-zag absorbing boundaries 
and have thus calculated the associated  absorption probabilities.
This problem was previously considered to be intractable \cite{KM63}.
We hope that our result will inspire further work in this area.

\ack

We thank Michael Barber for first drawing our attention to \cite{MW40} and
Wolfgang Schief for his warm oral translation of \cite{C28}.
MTB has been supported by The Australian Research Council.

\end{document}